\documentclass[aps,prb,twocolumn,showpacs,preprintnumbers,amsmath,amssymb,superscriptaddress]{revtex4}
\usepackage{graphicx}
\usepackage{dcolumn}
\usepackage{bm}
\usepackage{color}
\usepackage{ulem}

\newcommand{\labs} {\left\vert}
\newcommand{\rabs} {\right\vert}

\newcommand{\lab} {\left\langle}
\newcommand{\rab} {\right\rangle}

\begin{document}

\title{Manifestation of the shape and edge effects in spin-resolved transport
through graphene quantum dots}

\author{I. Weymann}
\email{weymann@amu.edu.pl} \affiliation{Department of Physics,
Adam Mickiewicz University, 61-614 Pozna\'n, Poland}

\author{J. Barna\'s}
\affiliation{Department of Physics, Adam Mickiewicz University,
61-614 Pozna\'n, Poland} \affiliation{Institute of Molecular
Physics, Polish Academy of Sciences, 60-179 Pozna\'n, Poland}

\author{S. Krompiewski}
\affiliation{Institute of Molecular Physics, Polish Academy of
Sciences, 60-179 Pozna\'n, Poland}

\date{\today}

\begin{abstract}
We report on theoretical studies of transport through graphene
quantum dots weakly coupled to external ferromagnetic leads. The
calculations are performed by exact diagonalization of a
tight-binding Hamiltonian with finite Coulomb correlations for
graphene sheet and by using the real-time diagrammatic technique
in the sequential and cotunneling regimes. The emphasis is put on
the role of graphene flake shape and spontaneous edge
magnetization in transport characteristics, such as the
differential conductance, tunneling magnetoresistance (TMR) and
the shot noise. It is shown that for certain shapes of the
graphene dots a negative differential conductance and nontrivial
behavior of the TMR effect can occur.
\end{abstract}

\pacs{73.63.Kv, 73.23.Hk, 73.22.Pr, 85.75.-d}

\maketitle


\section{Introduction}


Since its discovery,~\cite{novoselov2004} graphene has been
attracting an increasing attention due to its exceptional physical
properties and also possibilities of various promising practical
applications.~\cite{Geim2007,katsnelson,castro} For example, owing
to a very long spin diffusion length observed in graphene,~\cite{tombros07}
one may expect that graphene will
play an important role in future molecular spintronics. Moreover,
with the advent of new powerful experimental techniques, it is
possible now to engineer and fabricate graphene structures of
various shapes and sizes, ranging from sheets of large area to
extremely small graphene flakes. The latter can in particular exhibit
single-electron charging effects and, thus, behave as typical
quantum dots  - similar to quantum dots based on two-dimensional
electron gas. \cite{sols07,ponomarenko08,stampfer08,todd09}

In such small graphene flakes, the role of edges is much increased in
comparison to large graphene sheets. It is also well known on
theoretical grounds that zigzag edges of graphene nanostructures
have large densities of states, which in the presence of strong
enough on-site Coulomb repulsions can result in the appearance of
edge magnetism.~\cite{Fujita96,Yazyev10, Acik11} Indeed, it has
been confirmed experimentally by scanning tunneling spectroscopy
measurements that graphene nanoribbons and quantum dots reveal
highly enhanced densities of states (DOS) at the zigzag-type
fragments of their edges.~\cite{Klusek00, Kobayashi05} It is worth
noting, that the problem of graphene/graphite's edges has been recently
under intensive studies, as the carbon-based nanostructures can
potentially be used in modern nanoelectronics, including also
spintronics.~\cite{Sutter09,Son06,Kim08,Munoz-Rojas09, SK09,
Zhou10, Han11,Weymann08} Very recently, it has been
demonstrated experimentally that the edge DOSs in graphene
nanostructures are spin-split.~\cite{Tao11} Following this line,
in an attempt to gain additional insights into the edge states, we
suggest another approach to the problem, namely a visualization of
the effect of magnetic edges by the analysis of Coulomb blockade
spectra for graphene dots of different geometries.
To reach this objective we study the transport properties of
graphene quantum dots coupled to ferromagnetic leads.

As already mentioned above, in this paper we focus on the limit of
rather small graphene flakes, and address the transport properties
of graphene quantum dots weakly coupled to external ferromagnetic
leads. The  Coulomb blockade phenomena become then relevant. In
particular, we study the effects related with the shape and edges
of the graphene flakes on various spin-resolved transport
properties of the system, including differential conductance,
tunnel magnetoresistance (TMR) and shot noise (Fano factor).

The question whether or not edge states can be probed by
electronic transport methods is still a matter of intensive
discussion. On the one hand, the edge magnetism is critically
suppressed in the case of contacts of good (or even moderate)
transparency.~\cite{Areshkin12,MZ,SK} On the other hand, however,
the edge magnetism appears in isolated graphene flakes and also
survives when the flakes are weakly coupled to
electrodes.~\cite{Fernandez07,Yazyev10,Tao11} Additionally, the
edge states may be localized and therefore (very) weakly
conducting, which makes them hardly accessible by transport
measurements. Our studies show that some information on the edge
states can be extracted from transport measurements in the limit
of weak coupling between the graphene flakes and electrodes, and
when lateral dimensions of the flakes are not too large
(comparable to the localization length of the edge states). The
first assumption makes the energy spectra of the flakes rather
independent of the coupling to electrodes, while the second one
makes the edge states accessible in transport measurements (albeit
the corresponding conductance can be rather small). The Coulomb
blockade spectra provide then a kind of unique shape-specific
'fingerprints', which also contain some information on the edge
states.

The paper is organized as follows. First, the model as well as the
computational method based on real-time diagrammatic technique are
briefly outlined in Sec. II. Then, the numerical results are
presented and thoroughly discussed in Sec. III. Summary and final
conclusions are in Sec. IV.


\section{Theoretical framework}

\begin{figure}[t]
  \includegraphics[width=0.9\columnwidth]{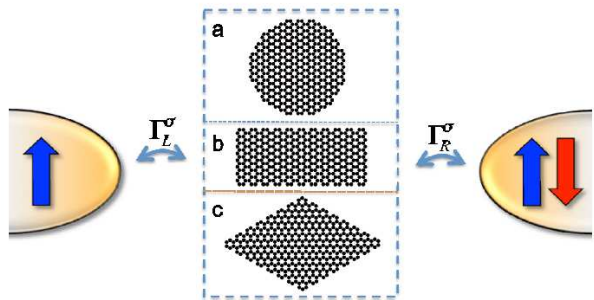}
  \caption{\label{Fig:1}
  (Color online) Scheme of a graphene quantum dot
  coupled to external ferromagnetic leads.
  We consider graphene flakes of three different shapes:
  circular (a), rectangular (b) and rhombic (c).
  A particular dot is coupled to external leads
  with coupling strengths described by $\Gamma_{L}^\sigma$ and
  $\Gamma_{R}^\sigma$ for the left and right leads.
  It is assumed that the system can be in two magnetic configurations:
  the parallel and antiparallel ones, as sketched in the figure.}
\end{figure}

The considered system is displayed in Fig.~\ref{Fig:1}. It
consists of a graphene flake that is weakly coupled to external
ferromagnetic leads. The coupling strengths are described by
$\Gamma_{L}^\sigma$ and $\Gamma_{R}^\sigma$ for the left and right
leads. We consider graphene flakes of three different hypothetical
shapes: circular, rectangular and rhombic. It is assumed that the
system can be in two magnetic configurations: either parallel or
antiparallel one, see Fig.~\ref{Fig:1}.

In order to find the energy levels as well as magnetic moments
of graphene quantum dots (GQDs) we have performed exact
diagonalization of the following mean-field Hamiltonian
\begin{equation} \label{H}
  H = - \sum \limits_{ i ,j, \sigma } t_{i,j}\labs i,
  \sigma \rab \lab \sigma, j \rabs +\frac{1}{2}\sum \limits_{i,
  \sigma} \Delta_{i, \sigma} \labs i, \sigma \rab \lab \sigma, i
  \rabs.
\end{equation}
Here, $t_{i,j}$ are the hopping integrals, $\labs i, \sigma \rab$
stand for $\pi$-electron orbitals at site $i$ with spin $\sigma$,
$ \Delta_{i, \sigma}=U(n_{i \sigma}-n_{i -\sigma})$ describes the
Stoner splitting, and $n_{i \sigma}$ are the respective occupation
numbers. The latter have been computed self-consistently by
summing up the squared eigenvectors corresponding to the
eigenvalues not greater than the Fermi energy. The hopping
integrals $t_{i,j}$ are assumed to be nonzero only for nearest
neighbors, and the nearest neighbor hopping parameter $t$ is set
to be equal $t=2.7$ eV. In turn, the Coulomb on-site repulsion is
assumed to be $U=1.2 t$ (see e.g. Ref.~[\onlinecite{Yazyev10}]).
All the GQDs we consider are of comparable area ($\sim$9 nm$^2$)
and consist of 350-400 carbon atoms.
Here we focus on the shapes with
a relatively small number of zigzag type edge atoms, and
consequently few quasi-degenerate edge states in the vicinity of
the Dirac point for U=0 (for triangular and hexagonal structures
with purely zigzag edges see Refs.~[\onlinecite{Fernandez07,Potasz12}]).

Exact diagonalization of the Hamiltonian (\ref{H}) yields the eigenvalues
$\varepsilon_{l\sigma}$ that have been then used as an input to the mean-field Hamiltonian
for the graphene quantum dot
\begin{equation} \label{HGQD}
  H_{\rm GQD} = \sum_{l,\sigma} \varepsilon_{l\sigma} n_{l\sigma}
  + \frac{E_C}{2} \left(N - n_0 \right)^2 \,.
\end{equation}
Here, $E_C$ is the phenomenological charging energy of the dot,
$\varepsilon_{l\sigma}$ is the energy of the dot's orbital
discrete level $l$ for spin $\sigma$, $n_{l\sigma} =
d^\dag_{l\sigma} d_{l\sigma}$ denotes the particle number operator
for the level $l$, $N=\sum_{l,\sigma} n_{l\sigma}$, and $n_0$ is
the number of electrons in an electrically neutral quantum dot.

The leads are modeled by Hamiltonian of noninteracting
quasi-particles
\begin{equation}
  H_{\rm Leads} = \sum_{\alpha=L,R} \sum_{{\bf k},\sigma} \varepsilon_{\alpha {\bf k} \sigma} c^\dag_{\alpha {\bf k}
  \sigma} c_{\alpha {\bf k} \sigma} \,,
\end{equation}
where $c^\dag_{\alpha {\bf k}\sigma}$ creates a spin-$\sigma$
electron with wave vector ${\bf k}$ in lead $\alpha$ and
$\varepsilon_{\alpha {\bf k}\sigma}$ is the corresponding energy.
In turn, tunneling processes between the dot and the leads are
described by
\begin{equation}
  H_{\rm Tun} = \sum_{\alpha=L,R} \sum_{{\bf k}, l, \sigma} v_{\alpha l} \left[ d^\dag_{l \sigma} c_{\alpha {\bf k}
  \sigma} + c^\dag_{\alpha {\bf k} \sigma} d_{l \sigma}\right],
\end{equation}
with $v_{\alpha l}$ denoting the hopping matrix element between
the dot level $l$ and the lead $\alpha$. The broadening of the
GQD's levels can be  described by $\Gamma_{\alpha l}^\sigma = 2\pi
\rho_{\alpha}^\sigma |v_{\alpha l}|^2$, where $\rho_\alpha^\sigma$
is the spin-dependent density of states in the lead $\alpha$ for
spin subband $\sigma$. In the case of ferromagnetic leads this can
be then written as, $\Gamma_{\alpha l}^{+(-)} = (1\pm
p_{\alpha})\Gamma_{\alpha l}$, with $\Gamma_{\alpha l}=
(\Gamma_{\alpha l}^{+} +\Gamma_{\alpha l}^{-})/2$ and $p_\alpha$
being the spin polarization of lead $\alpha$. Here,
$\Gamma_{\alpha l}^{+}$ ($\Gamma_{\alpha l}^{-}$) corresponds to
the coupling to the majority (minority) spin band. In the
following we assume, $\Gamma_{\alpha l} = \Gamma_{\alpha} \equiv
\Gamma/2$ and $p_{\alpha } \equiv p$. Furthermore, in numerical
calculations, for all three different shapes of the dots, we have
also assumed the charging energy $E_C = 0.15$ eV and the coupling
strength $\Gamma = 0.002$ eV. The $E_C$ value has been found from
a scaling law (against QD-size) formulated in
Ref.~[\onlinecite{Ma09}]. Incidentally, this scaling leads also to
acceptable estimations of charging energies in
Ref.~[\onlinecite{ponomarenko08,stampfer08}].

In order to reliably determine the transport properties of Coulomb-blockade graphene
quantum dots weakly coupled to external leads, we employ the real-time
diagrammatic technique.~\cite{diagrams,thielmann,weymannPRB08}
This technique relies on systematic perturbation expansion of the
reduced density matrix and the operators of interest in the dot-lead coupling strength $\Gamma$.
The calculation proceeds with the determination of respective self-energy matrices
$\mathbf{W}$, which enables the evaluation of the elements of the reduced density matrix of GQD
by using the following master-like equation~\cite{thielmann}
\begin{equation}\label{Eq:master}
  (\mathbf{\tilde{W}}\mathbf{p}^{\rm st})_{\chi} =
  \Gamma\delta_{\chi\chi_0}\,.
\end{equation}
Here, $\mathbf{p}^{\rm st}$ is the vector containing stationary
probabilities, $\mathbf{\tilde{W}}$ is the modified self-energy
matrix $\mathbf{W}$ so as to include the normalization of
probabilities and $\chi\equiv l\sigma$ labels the states of the
graphene quantum dot. Having found the occupation probabilities,
the current flowing through the system can be calculated using the
following equation~\cite{thielmann}
\begin{equation}\label{Eq:current}
  I=\frac{e}{2\hbar}{\rm Tr}\{\mathbf{W}^{\rm I}\mathbf{p}^{\rm st}\}\,,
\end{equation}
where $\mathbf{W}^{\rm I}$ is the modified self-energy matrix $\mathbf{W}$
to account for the number of electrons transferred through the system.

By performing the perturbation expansion of the respective
self-energies one is then able to calculate the current order by
order in tunneling processes. In this paper we have included the
first and second-order self-energies. The first order of expansion
corresponds to sequential tunneling, which dominates transport
outside the Coulomb blockade regime, whereas the second-order
self-energies describe the cotunneling processes. Cotunneling
processes occur through virtual states of the system and are
dominant in the Coulomb blockade regime.~\cite{cotunneling} Thus,
to properly describe the transport properties of the system in the
full range of bias and gate voltages, it is of vital importance to
include both the first and second order terms of the perturbation
expansion.

With the aid of the real-time diagrammatic technique, we have
determined the behavior of the current $I$, differential
conductance $G$ and tunnel magnetoresistance TMR of graphene
quantum dots in both the linear and nonlinear response regimes. In
addition, we have also calculated the bias and gate voltage
dependence of the shot noise $S$ and the corresponding Fano factor
$F$.~\cite{thielmann} The Fano factor is defined as, $F =
S/(2e|I|)$, and describes deviation of the shot noise from the
Poissonian value, relevant for uncorrelated tunneling
events.~\cite{blanterPR00} The main results and their discussion
are presented in the sequel.


\section{Results and discussion}


Below we present numerical results on transport through graphene
quantum dots of similar sizes (in terms of the number of atoms),
but different shapes. For simplicity, we assume the Fermi level of
the leads to be equal to the Fermi level of a neutral graphene. We
show that the corresponding energy spectra are strongly dependent
on the GQD geometry, and that the emerging magnetic moments are
essentially localized at zigzag-like segments of the dots' edges.
This leads to different transport characteristics of particular
GQDs, as shown and discussed in the following. More specifically,
we show that the size of blockade regions (blockade diamonds in
the bias-gate voltage dependence of the differential conductance)
can be used to gain some information about the edge states. As it is
well known, the size of the diamonds is determined by the Coulomb
charging energy $E_C$ and the level spacing. Therefore, the
sequence of diamonds depends on whether a given level is spin
degenerate or not. The absence of spin degeneracy, in turn,
implies the presence of magnetic states. Since the edge states are
either in the center of the energy gap or close to it, the
corresponding diamonds can be easily identified.

\begin{figure}[t]
  \includegraphics[width=0.8\columnwidth]{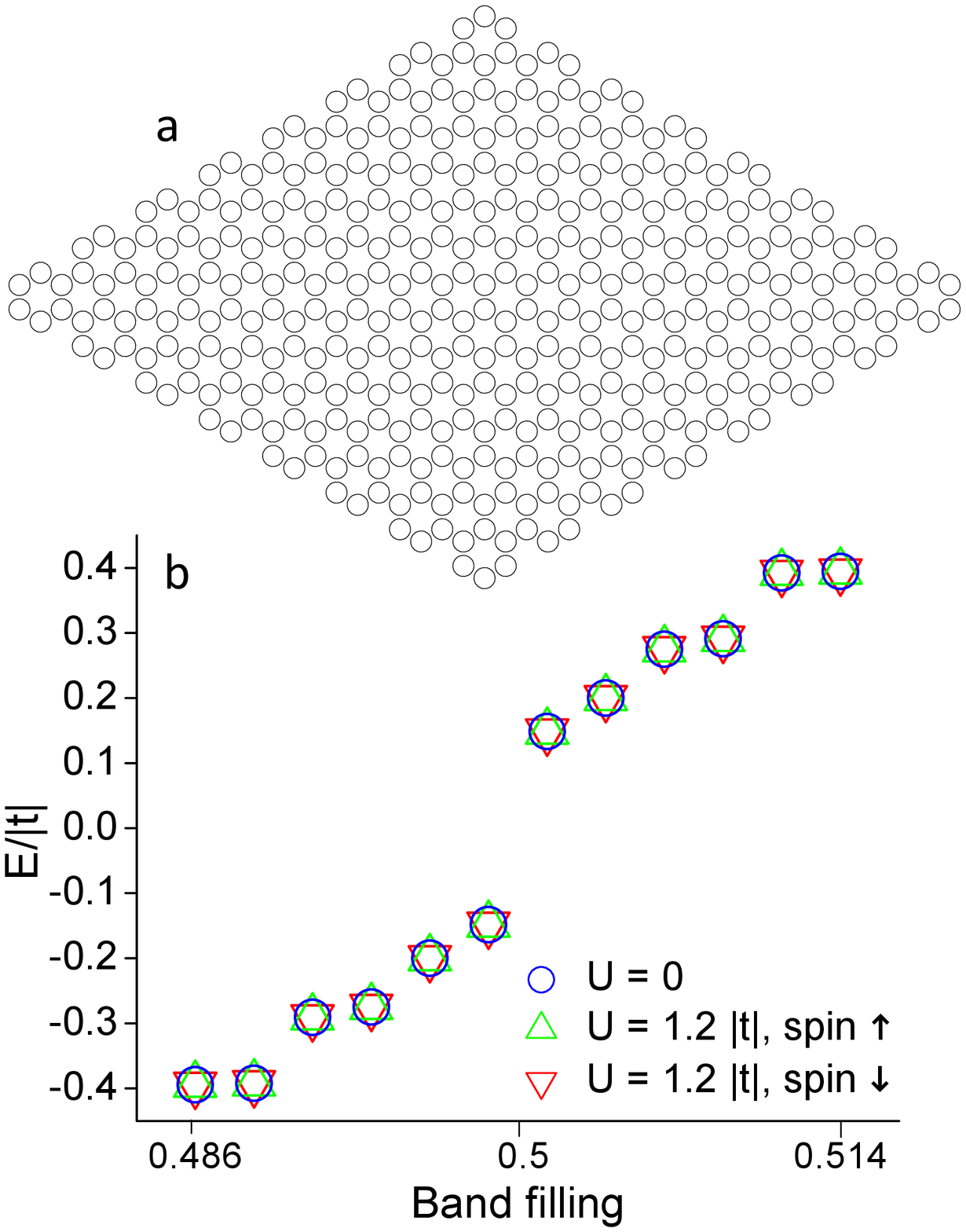}
  \caption{\label{Fig:2}
  (Color online) The atomic structure (a) and
  energy spectrum (b) of the rhombic graphene quantum dot.
  There are no magnetic solutions irrespective of the value of $U$,
  instead a quite pronounced energy gap opens.}
\end{figure}


\subsection{Rhombic graphene dots}


Let us begin our considerations with the case of graphene flake of
rhombic geometry with exclusively armchair-type edges. The atomic
structure of the graphene dot and the corresponding energy
spectrum measured from the Fermi level, $E_F=0$, is shown in
Fig.~\ref{Fig:2}. As one can see in this figure, the rhombic
graphene flake has neither low-energy localized states nor
magnetic moments at the edges. Moreover, the energy levels of the
dot are independent of the Coulomb parameter $U$, which is a
consequence of the absence of magnetized states. As one can
readily see, such a structure has therefore a quite pronounced
energy gap at the Fermi level. Thus, the armchair-edge rhombic
geometry may be suitable for engineering graphene nanostructures
useful for field effect transistor devices (with a pronounced
ON/OFF current ratio).

\begin{figure}[t]
  \includegraphics[width=0.95\columnwidth]{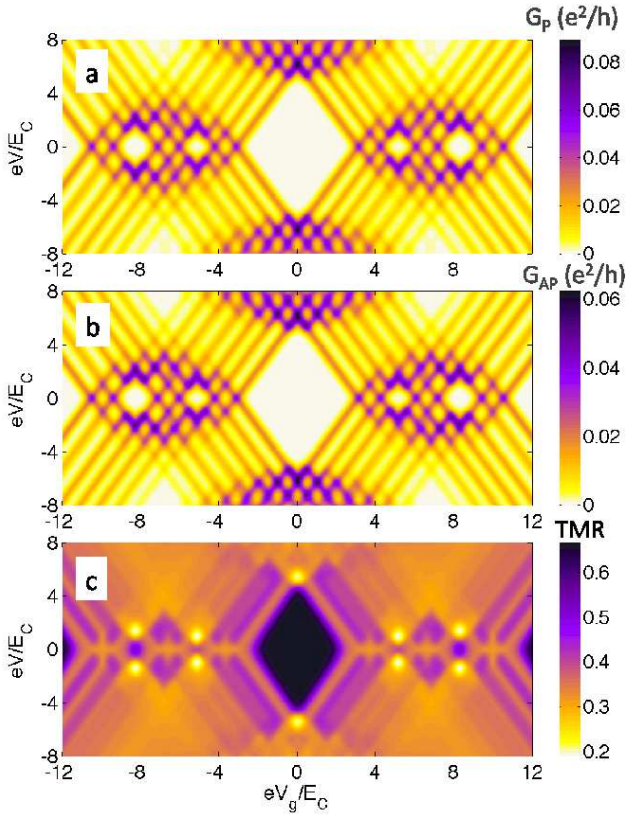}
  \caption{\label{Fig:3}
  (Color online) The bias $V$ and gate $V_g$ voltage dependence of
  the differential conductance in the parallel (a) and antiparallel (b) configurations
  as well as the resulting TMR effect (c) calculated for rhombic-like graphene quantum dot.
  The parameters are: the charging energy $E_C = 0.15$ eV, the coupling strength $\Gamma = 2$ meV,
  the thermal energy $k_BT = 20$ meV and the leads' spin polarization $p=0.5$.}
\end{figure}

The bias and gate voltage dependence of the differential
conductance in the parallel and antiparallel configurations is
shown in Figs.~\ref{Fig:3}(a) and \ref{Fig:3}(b). The central
white region corresponds to zero excess electrons in the dot. This
region  is relatively large due to the large  energy gap. With
sweeping the gate voltage, one shifts the position of the graphene
dot levels, changing thus the number of electrons in the dot. Due
to particle-hole symmetry the spectrum is symmetric with respect
to the central diamond. The second diamond (with increasing gate
voltage  starting from $V_g=0$) is much smaller, when compared to
the middle one, as it is determined only by the charging energy
$E_C$. The next diamond, apart from the charging energy, also
includes the level spacing between the first and second levels
(for positive or negative energies). Since each level is spin
degenerate, every second diamond
(to the left or to the right of the central diamond) is determined only by the charging energy
$E_C$ and therefore all of them are relatively small. The other
diamonds include additionally the level spacing and are therefore
larger. Thus, the level spacings and the energy gap can, in
principle, be determined from the size of the corresponding
blockade diamonds.

\begin{figure}[t]
  \includegraphics[width=0.95\columnwidth]{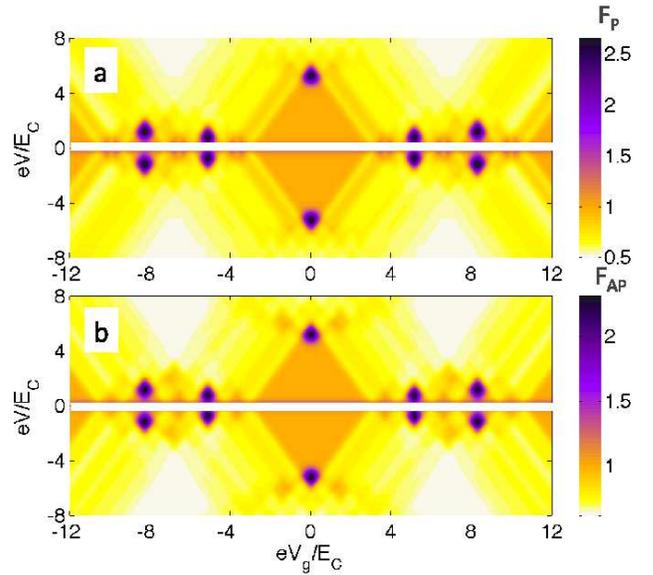}
  \caption{\label{Fig:4}
  (Color online) The Fano factor in the parallel (a) and antiparallel (b)
  magnetic configuration as a function of the bias and gate voltages
  for rhombic-like graphene quantum dot. The parameters are the same as in Fig.~\ref{Fig:3}.}
\end{figure}

Except for particular structure of the Coulomb diamonds related
with the energy spectrum of the dot, which is independent of the
magnetic configuration of the device, one can see that the
differential conductance in the parallel configuration $G_{\rm P}$
is larger than that in the antiparallel configuration $G_{\rm
AP}$. This is related with the asymmetry in the couplings to the
spin majority and spin minority bands of the ferromagnets. For the
parallel configuration, the majority-majority channel is most
conducting, while in the antiparallel configuration there are two
weakly conducting majority-minority channels. As a consequence,
$G_{\rm P} > G_{\rm AP}$. The difference between the system
transport properties in these two magnetic configurations is
described by the tunnel magnetoresistance, defined as,~\cite{julliere} ${\rm TMR}
= (R_{\rm AP}-R_{\rm P})/R_{\rm P}$, with $R_{\rm P}$ and $R_{\rm
AP}$ denoting the resistance in the parallel and antiparallel
magnetic configurations of the device. The bias and gate voltage
dependence of TMR is shown in Fig.~\ref{Fig:3}(c). First of all,
one can see that the TMR is always positive, as in typical
spin-value quantum dot devices.~\cite{weymannPRB05} The TMR is
particularly enhanced in the central Coulomb blockade diamond
[black area in Fig.~\ref{Fig:3}(c)], where the sequential
tunneling is exponentially suppressed while transport takes place
{\it via} cotunneling events. In this transport regime the dot is
empty and the current is driven by elastic cotunneling
processes.~\cite{weymannPRB05,weymannPRB08} The TMR is then
exactly given by the Julliere value,~\cite{julliere}
$\rm{TMR^{Jull}}=2p^2/(1-p^2)$, which for the assumed parameters
yields $2/3$. In other Coulomb diamonds, the inelastic cotunneling
processes become relevant and the TMR is generally smaller than
Julliere's value. Similar behavior can be observed in the
sequential tunneling regime, where the TMR is much smaller than
${\rm TMR^{Jull}}$. This is related with the fact that the
information about magnetic configuration of the device is now
transferred by uncorrelated sequential tunneling events and the
non-equilibrium spin accumulation which builds up in the graphene
quantum dot. This mechanism is less effective than direct
spin-conserving cotunneling between the left and right lead, thus
the TMR is smaller than Julliere's value.

The Fano factors in the parallel and antiparallel
magnetic configurations of the device are shown in
Fig.~\ref{Fig:4}. In both cases, the shot noise in the blockade
regions, where cotunneling processes dominate, is generally
super-Poissonian, i.e. the corresponding Fano factors are larger
than unity, $F\gtrsim 1$. This is related with bunching of
inelastic cotunneling events~\cite{sukhorukovPRB01} and has
already been observed experimentally in transport through other quantum dot
systems.~\cite{onac06,zhang} Interestingly, in the central
diamond, where elastic cotunneling mediates the current, the Fano
factors in both magnetic configurations approach unity. This is
due to the fact that the elastic cotunneling processes are
uncorrelated in time and the shot noise is Poissonian. In the
regions, where transport is dominated by sequential tunneling
processes, the corresponding shot noise is rather sub-Poissonian
with the relevant Fano factors smaller than 1, $F<1$. This
suppression of the noise is due to Coulomb correlations between
consecutive sequential tunneling processes. We note, that the
calculations include also the thermal noise, which is dominant in
the small bias voltage regime. Accordingly, when $V$ tends to
zero, the current and shot noise both tend to zero, and
the noise is dominated by the thermal noise (nonzero also at
$V=0$). Thus, the corresponding Fano factor diverges when $V\to 0$,
as marked with a horizontal stripe in Fig.~\ref{Fig:4}.


\subsection{Circular graphene dots}


\begin{figure}[t]
  \includegraphics[width=0.8\columnwidth]{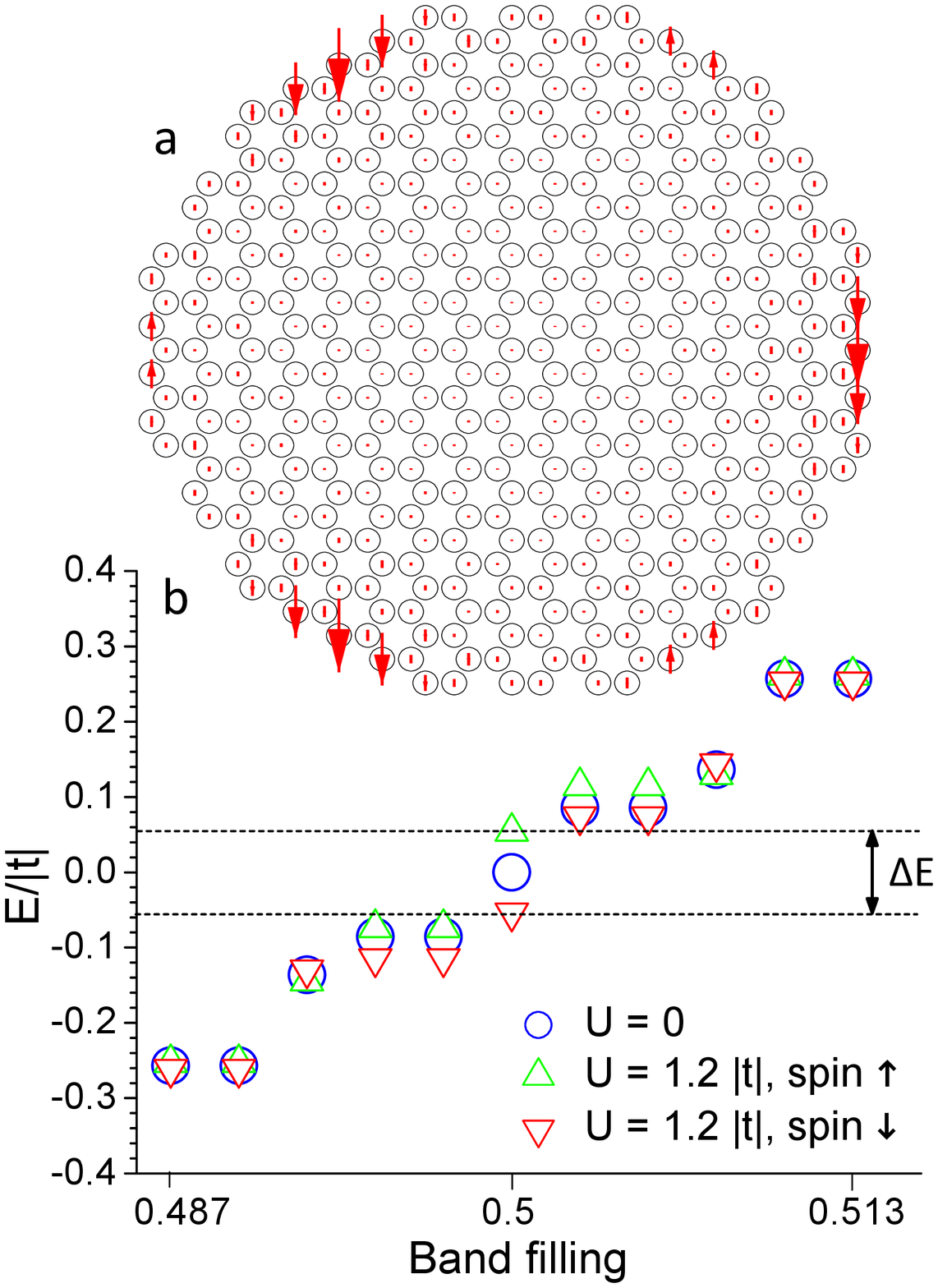}
  \caption{\label{Fig:5}
  (Color online) The atomic structure (a)
  and energy spectrum  (b)
  for circular graphene flake.
  In the case of $U=0$, there are no magnetic edges,
  while magnetic edges appear for finite $U$.
  Note that in the latter case an
  energy gap ($\Delta E$) opens. The corresponding magnetic
  configuration is also displayed in (a).
  The maximum edge magnetic moments are equal to ca. 1/5 $\mu_B$.
  However, the total magnetic moment of the dot is equal to $1\mu_B$.}
\end{figure}

Let us now have a closer look at the circular graphene flake. In
this geometry the edge atoms have rather short zigzag-like (and
armchair-like) coordinations. Consequently, any complete
compensation of the edge magnetization (antiferromagnetic
alignment) can be hardly realized on geometrical grounds. Our
calculations show that indeed the ground state magnetic configuration
of the edges is ferromagnetic-like (ferrimagnetic), but with some admixture of
antiparallel magnetic moments. In this situation there is no way
to control spin dependent electric current with external magnetic
field, unless one makes use of ferromagnetic electrodes.

When $U=0$ (nonmagnetic dot), there is no gap at the Fermi level.
However, as shown in Fig.~\ref{Fig:5}, the presence of edge
magnetism ($U>0$) leads to the opening of an energy gap $\Delta E$
at the Fermi level. Noteworthy, near the half-band filling the
highest occupied energy levels are fully spin-polarized - the dot
behaves as a magnetic one. The overall magnetic moment of the dot
shown in Fig.~\ref{Fig:5} is equal to $1\mu_B$. This, in turn,
affects the spin-resolved transport properties of the system,
which now show some asymmetry with respect to the bias and gate
voltage reversal.

\begin{figure}[t]
  \includegraphics[width=0.95\columnwidth]{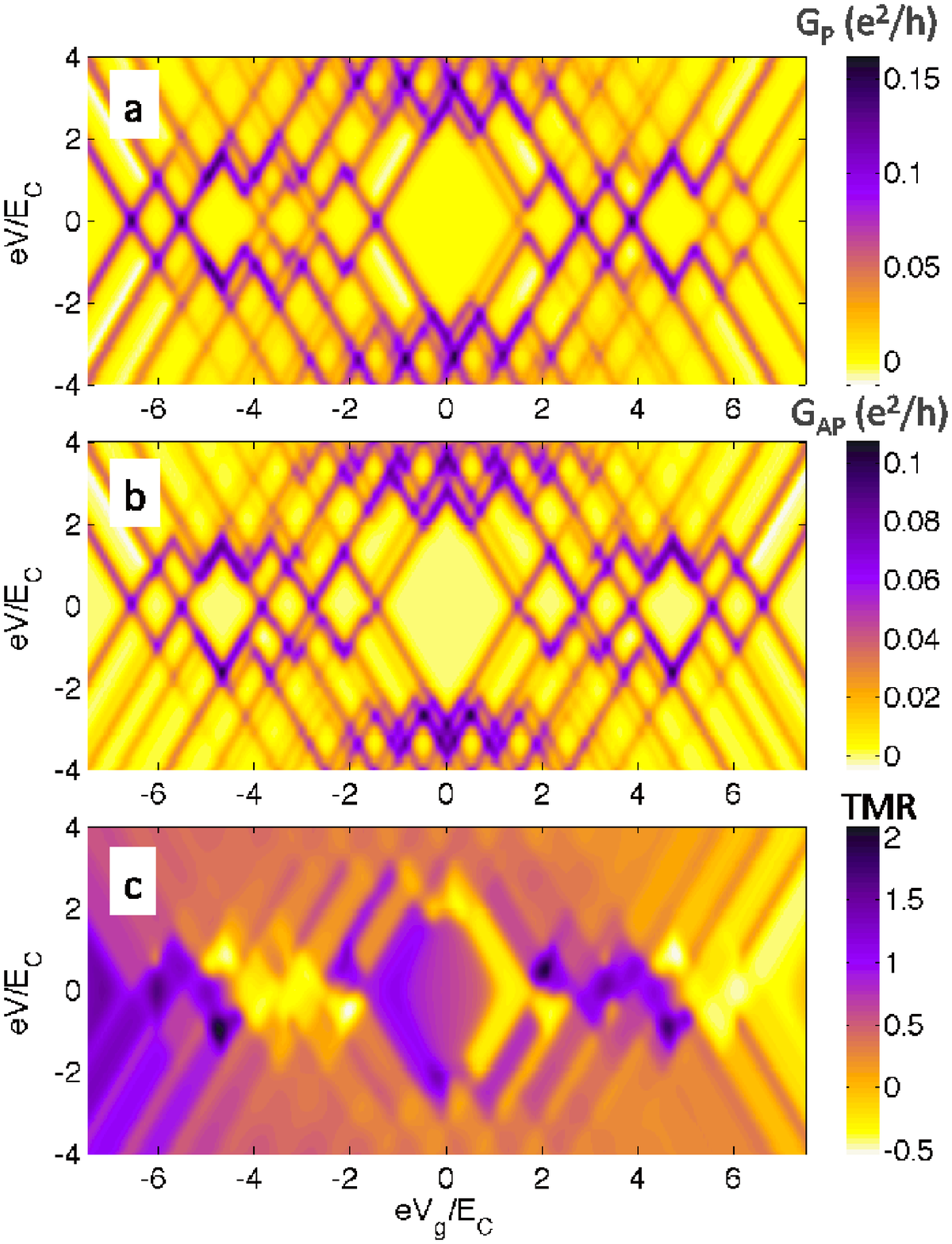}
  \caption{\label{Fig:6}
  (Color online) The bias and gate voltage dependence of
  the differential conductance in the parallel (a) and antiparallel (b) configurations
  as well as the resulting TMR effect (c) calculated for circular graphene quantum dot.
  The parameters are the same as in Fig.~\ref{Fig:3} except
  for thermal energy which is $k_BT = 10$ meV.}
\end{figure}

Incidentally, in accordance with Lieb's theorem,~\cite{lieb89}
one can increase the net magnetic moment up to
$(N_A-N_B)\mu_B$, by modifying  the edges so as to increase the
imbalance of the numbers $N_A$ and $N_B$ of the A(B)-sublattice
atoms. Our results are rather robust against moderate edge
disorder. In particular, when a few edge atoms are removed then
possible implications are consistent with predictions based on the
aforementioned theorem. Noteworthy, the asymmetry seen in Fig.~\ref{Fig:6}
provides direct visualization of the presence of GQD's
uncompensated spin. In another context as the present one, the
edge state effects on the electronic structure as well as charge
and spin transport have been studied in Refs.~[\onlinecite{Libisch09,Wimmer08}].

\begin{figure}[t]
  \includegraphics[width=0.95\columnwidth]{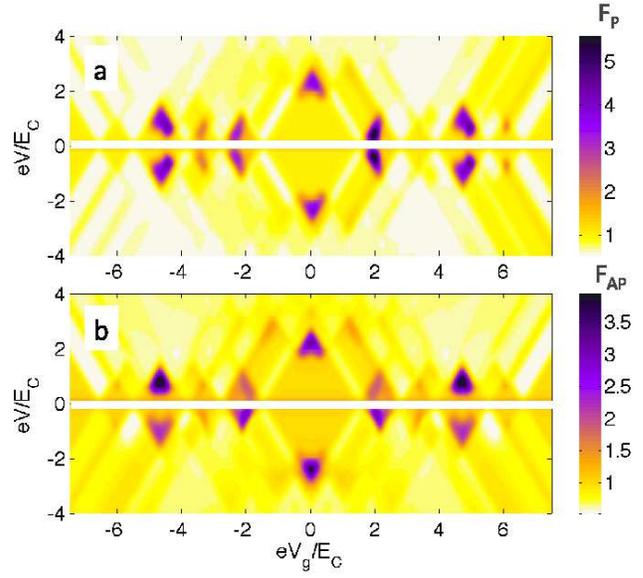}
  \caption{\label{Fig:7}
  (Color online) The Fano factor in the parallel (a) and antiparallel (b)
  magnetic configuration as a function of the bias and gate voltages
  for circular graphene quantum dot. The parameters are the same as in Fig.~\ref{Fig:6}.}
\end{figure}

The differential conductance spectra in both magnetic
configurations are shown in Fig.~\ref{Fig:6}. First, we note that
in equilibrium all the states of negative energy are occupied and
the dot has one unpaired spin-down electron. The addition of a new
electron costs then the charging energy plus half the level
spacing $\Delta E$, see Fig.~\ref{Fig:5}. Note that the added
electron is a spin-up one. This corresponds to the large central
diamond in Fig.~\ref{Fig:6}. The subsequent electron occupies
spin-down level and the corresponding addition energy includes the
Coulomb charging energy plus the spacing between the first and
second levels of positive energy. Since the latter spacing is mach
smaller than $\Delta E$, the corresponding diamond is smaller than
the central one, but it is still larger than the diamond
determined by $E_C$ only. It indicates that this state is
magnetic. The next electron occupies again the spin-down level,
which is almost degenerate with the previous level. The
corresponding diamond is determined practically by the charging
energy $E_C$ and is therefore smaller than the preceding one. The
following two electrons occupy the two spin-up states (also almost
degenerate), so the corresponding two diamonds are determined by
the charging energy plus the level spacing and charging energy,
respectively. Similar scenario holds for higher gate voltages,
when higher energy levels become occupied by electrons, as well as
to negative gate voltages.

It is also worth noting that now the spectra are significantly
different from the corresponding ones for $U=0$, see
Fig.~\ref{Fig:5}. Since for $U=0$ there is no gap at the Fermi
level, the central diamond should be then determined by the
charging energy only and therefore should be small. The second
diamond should be relatively large while the three subsequent
diamonds should be small and determined by the charging energy.
Thus, the conductance spectra can be used to distinguish the
situation with nonzero $U$ from that with $U=0$.

Apart from the typical Coulomb blockade diamonds described above,
the transport characteristics display an asymmetry with respect
to the bias reversal, see Fig.~\ref{Fig:6}. This is due to the spin splitting
of the GQD's levels. In addition,
in certain transport regimes one can find a negative differential
conductance, which is present in both magnetic configurations.
This effect is basically related with the fact that for certain
transport voltages electrons participating in transport are mainly
spin-down (minority-spin) ones and the current is thus decreased.

The difference between the currents flowing through the system in
the two magnetic configurations of the device results in the TMR
effect shown in Fig.~\ref{Fig:6}c. Because now the energy
eigenstates of the graphene dot are no longer spin degenerate, one
can observe a nontrivial behavior of the TMR as a function of both
the bias and gate voltages. First of all, it can be seen that in
certain transport regimes the TMR can take values much larger than
those given by Julliere's model. Furthermore, there are transport
regimes where TMR changes sign and becomes negative. To explain
this behavior we will refer to approximate formulas for sequential
tunneling and cotunneling currents. Suppose the temperature is
much larger than the coupling strength $\Gamma$, but still smaller
than the charging energy $E_C$, $E_C > k_BT \gg \Gamma$. Then,
transport is mainly determined by thermally-activated sequential
tunneling. The current is thus proportional to, $I^{\rm seq} \sim
\sum_\sigma \Gamma_{L}^\sigma \Gamma_{R}^\sigma /
(\Gamma_{L}^\sigma + \Gamma_{R}^\sigma)$. If transport occurs
through the spin-up level of the dot, the current in the parallel
configuration is, $I_{\rm P}^{\rm seq} \sim (1+p)\Gamma/2 $, while
for the antiparallel configuration one finds, $I_{\rm AP}^{\rm
seq} \sim (1-p^2) \Gamma/2$, which yields for the TMR, ${\rm
TMR^{seq}} = p/(1-p)$. If, in turn,  the current is related with
tunneling through spin-down levels, one gets $I_{\rm P}^{\rm seq}
\sim (1-p)\Gamma/2 $ and $I_{\rm AP}^{\rm seq} \sim
(1-p^2)\Gamma/2$ for the parallel and antiparallel configurations,
respectively, and the TMR is now negative, ${\rm TMR^{seq}} =
-p/(1+p)$. On the other hand, if the temperature is comparable to
the coupling strength, $k_BT\approx \Gamma \ll E_C$, the current
is mainly due to cotunneling processes. Its dependence on the spin
polarization of the leads can be found if one considers elastic
cotunneling regime, where $I\sim \sum_\sigma \Gamma_L^\sigma
\Gamma_R^\sigma$. Then, if cotunneling occurs through spin-up
levels, the currents in the two configurations are proportional
to, $I_{\rm P} \sim (1+p)^2\Gamma^2$, $I_{\rm AP} \sim
(1-p^2)\Gamma^2$, yielding ${\rm TMR} = 2p/(1-p)$, which for the
assumed parameters ($p=0.5$) gives, ${\rm TMR} = 2$. Note that now
${\rm TMR} = 3 \times {\rm TMR^{Jull}}$, i.e. one obtains maximum
TMR three times larger than in the case of rhombic-like graphene
flake discussed in the  previous section. However, if cotunneling
occurs through spin-down levels, the current in the antiparallel
configuration is the same as above, while the current in the
parallel configuration becomes, $I_{\rm P} \sim (1-p)^2\Gamma^2$,
which results in  ${\rm TMR} = -2p/(1+p)$. For the assumed
parameters one then finds, ${\rm TMR} = -2/3 = -{\rm TMR^{Jull}}$,
i.e. the system exhibits a large-magnitude negative TMR. Because
the calculations are performed for $E_C > k_BT > \Gamma$, one
finds that the TMR in the Coulomb diamonds is much enhanced
compared to Julliere's value, with  $2p/(1-p) \gtrsim {\rm TMR}
\gtrsim p/(1-p)$, if transport is due to spin-up states of the
graphene dot. On the other hand, if spin-down states are active in
transport a negative TMR occurs with, $-p/(1+p) \lesssim {\rm TMR}
\lesssim -2p/(1+p)$. These values of the TMR can be clearly seen
in Fig.~\ref{Fig:6}c.

\begin{figure}[t]
  \includegraphics[width=0.8\columnwidth]{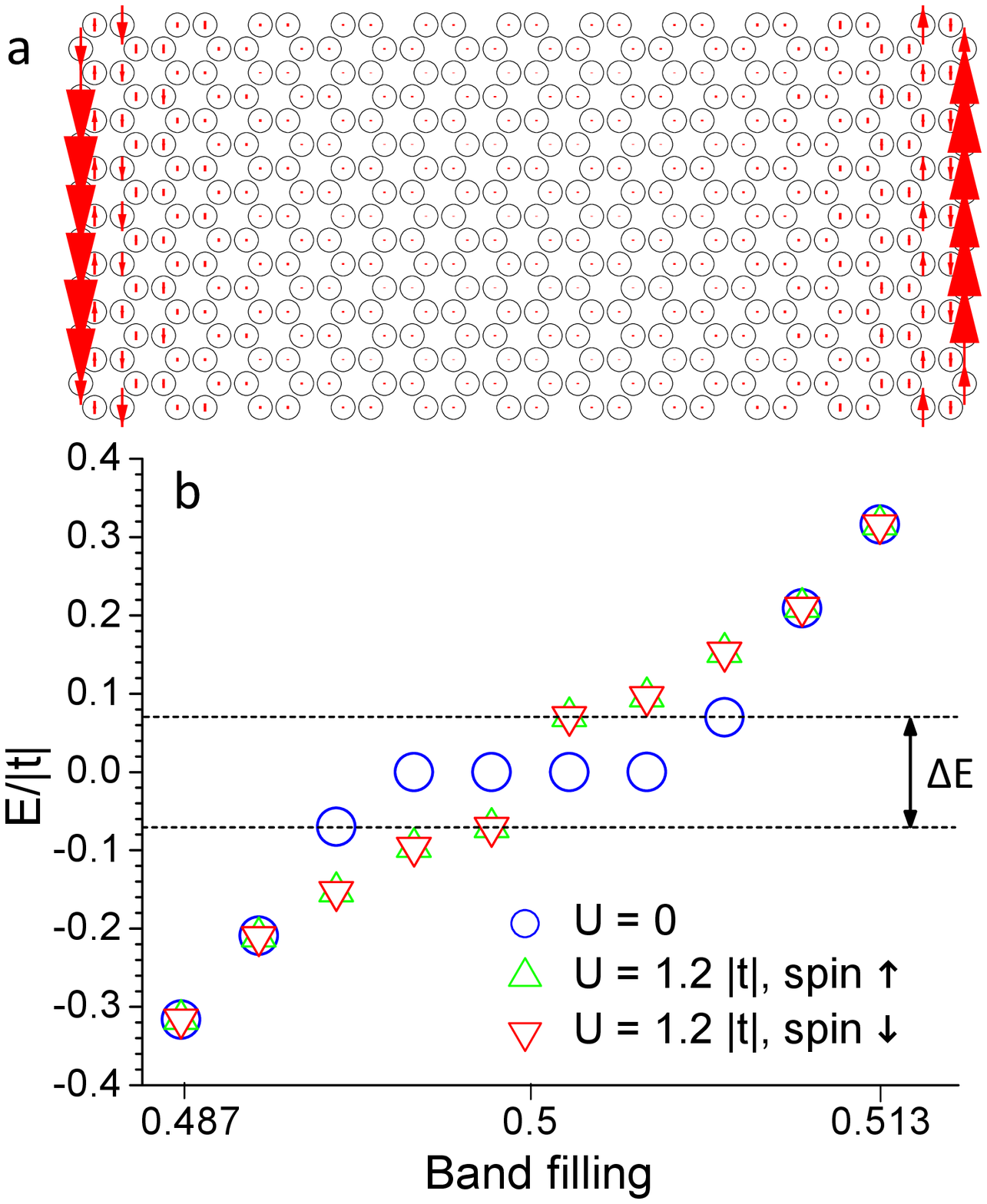}
  \caption{\label{Fig:8}
  (Color online) The same as in  Fig.~\ref{Fig:5} but for the rectangular
  graphene quantum dot. Now the maximum edge magnetic
  moments are equal to ca. 1/3 $\mu_B$.}
\end{figure}

The corresponding Fano factors in both magnetic configurations are
shown in Fig.~\ref{Fig:7}. In both cases, the shot noise is
generally super-Poissonian in the Coulomb blockade regions, where
cotunneling processes dominate, and bunching of inelastic events
enhances the noise. In other regimes transport is dominated by
sequential tunneling processes and the corresponding shot noise is
sub-Poissonian, $F<1$. As before, this is due to the Coulomb
correlations between sequential tunneling events and the Pauli
exclusion principle, resulting in suppressing of the noise.


\subsection{Rectangular graphene dots}


\begin{figure}[t]
  \includegraphics[width=0.95\columnwidth]{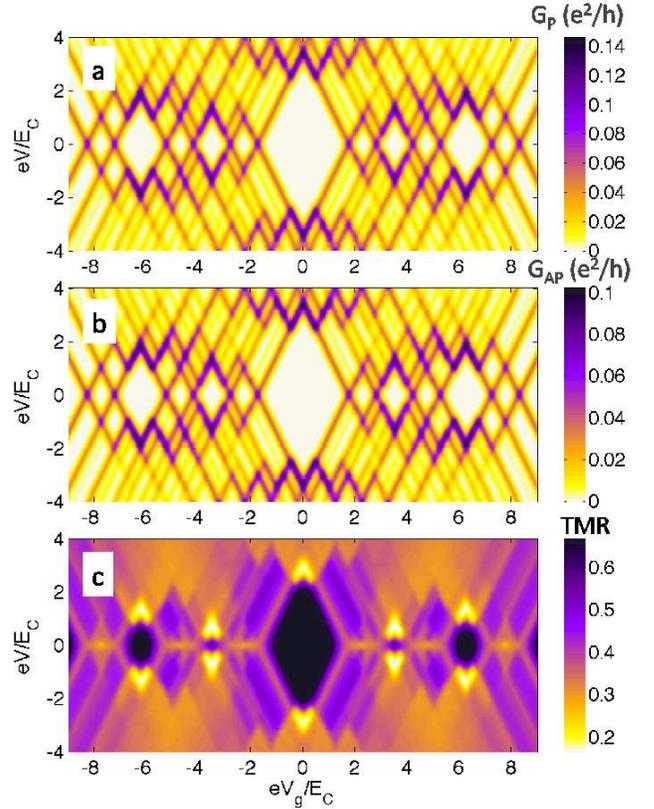}
  \caption{\label{Fig:9}
  (Color online) The bias and gate voltage dependence of
  the differential conductance in the parallel (a) and antiparallel (b) configurations
  as well as the resulting TMR effect (c) calculated for rectangular graphene quantum dot.
  The parameters are the same as in Fig.~\ref{Fig:6}.}
\end{figure}

\begin{figure}[t]
  \includegraphics[width=0.95\columnwidth]{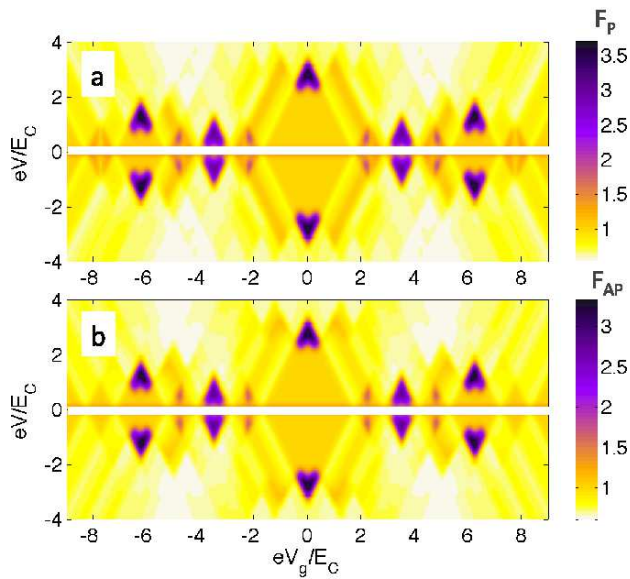}
  \caption{\label{Fig:10}
  (Color online) The Fano factor in the parallel (a) and antiparallel (b)
  magnetic configuration as a function of the bias and gate voltages
  for rectangular graphene quantum dot.
  The parameters are the same as in Fig.~\ref{Fig:6}.}
\end{figure}

In this subsection, transport through a rectangular graphene
quantum dot with zigzag and armchair edges is studied. In contrast
to the cases discussed earlier, with either nonmagnetic edges
(rhombus dot) or nearly ferromagnetically aligned ones (circular
dot), the present case corresponds to a spontaneous antiparallel
alignment of magnetic edges. This device may therefore act as an
effective spin-valve or a magnetoresistive sensor, and appears
potentially useful for spintronics.~\cite{Son06,Maassen11}

Figure~\ref{Fig:8} presents  atomic structure of the considered
graphene flake and the corresponding energy spectrum. It can bee
seen that for a sufficiently large on-site Coulomb repulsion
parameter $U$, an energy gap opens in the energy spectrum,
resulting in a metal/semiconductor (zero-gap/finite-gap
semiconductor) transition.
Moreover, the energy levels of this graphene dot remain spin
degenerate (perfect compensation of the net magnetization),
in contrast to the case of the circular dot.

The subsequent figures show the corresponding bias and gate
voltage dependence of the differential conductance and TMR, see
Fig.~\ref{Fig:9}, as well as the Fano factor, see
Fig.~\ref{Fig:10}. Since there is a gap at the Fermi level, the
central diamond in the differential conductance is now rather
large and includes the charging energy plus a half of the energy
gap. Because the energy levels are now spin degenerate, every
second diamond is smaller, as it is determined merely by the
charging energy $E_C$. Moreover, because of the spin degeneracy of
the levels, the differential conductance in the parallel
configuration is always larger than that in the antiparallel
configuration, yielding positive TMR, see Fig.~\ref{Fig:9}. The
behavior of TMR resembles that in the case of the rhombic graphene
flake. The TMR is given by Julliere's value in the Coulomb
diamonds where elastic cotunneling drives the current, and is much
more suppressed in the other diamonds with either inelastic
cotunneling or sequential tunneling.

The bias and gate voltage dependence of the Fano factor in both
magnetic configurations of the device is shown in Fig.~\ref{Fig:10}. The general
features of these spectra are qualitatively similar to those
discussed earlier in the case of the rhombic-like graphene dots:
$F \approx 1$ in the Coulomb blockade regions when elastic
cotunneling is relevant, $F\gtrsim 1$ in the blockade regions when
inelastic cotunneling is present, and $F<1$ in the sequential
tunneling regime.


\section{Conclusions}


In this paper we have analyzed transport properties of graphene
quantum dots of different geometries in the presence of the
on-site Coulomb repulsion. Characteristic features due to GQD
edges are shown to be reflected in the Coulomb blockade diamonds
and other transport characteristics. It has been demonstrated
that the TMR effect in graphene quantum dots may be quite large,
much larger then the corresponding Julliere's value of TMR. In
turn, the shot noise was found to take both sub-Poissonian as well
as super-Poissonian values. Out of the three graphene flake shapes
studied here, the armchair-edge rhombus flake shows no magnetism,
the circular flake has uncompensated ferrimagnetic edge
configuration, whereas the rectangular flake displays
antiparallelly aligned zigzag-edge magnetizations. All these
features manifest themselves in transport characteristics,
providing information about the geometry and edge states of
graphene dots.

In frame of the model assumed, the conductance spectra can be used
to check whether the energy spectrum includes zero energy levels
or not. If they do, then the first diamonds are determined only by
the charging energy and are small. If not, then the central
diamonds are large as they include not only the charging energy,
but also the level separation due to the size quantization.
However, the model is not free from some limitations, so the
conductance spectra in real situations may be different.
Accordingly, the conclusions concerning magnetic edge states
should rather be regarded as indicative only. First of all, the
description is based on mean-field approximation for charging
energy $E_C$ and also for the coupling parameter $\Gamma$. In
reality, both parameters may depend on the particular states
taking part in transport. As already mentioned in the
introduction, this may be especially important for $\Gamma$, when
transport occurs {\it via} localized edge states. As the variation
of $E_C$ has some influence on the size of the Coulomb diamonds,
the variation of $\Gamma$ modifies mainly the conductance and TMR,
leaving the size of the diamonds unchanged. Being aware of all the
underlying assumptions, we believe our results present a sound
starting point aimed at transport characterization of the
electronic states (including also the edge states) of graphene
flakes. Since both the crucial parameters (especially the coupling
parameter $\Gamma$) depend on a particular experimental
realization, a more accurate description can be performed when the
relevant experimental data are available.


This work was supported by the Polish Ministry of Science and
Higher Education as a research project No. N N202 199239 for years 2010-2013.
I.W. also acknowledges support from the Alexander von Humboldt Foundation
and the 7FP of the EU under REA grant agreement No CIG-303 689.



\begin{thebibliography}{99}


\bibitem{novoselov2004}
K.S. Novoselov, A.K. Geim, S.V. Morozov, D. Jiang, Y. Zhang, S.V.
Dubonos, I.V. Grigorieva, A.A. Firsov, Science {\bf 306}, 666 (2004).

\bibitem{Geim2007}
A. K. Geim and K. S. Novoselov, Nature Mater. \textbf{6}, 183
(2007).

\bibitem{katsnelson}
M. I. Katsnelson, Mater. Today \textbf{10}, 20 (2007).

\bibitem{castro}
A. H. Castro Neto, F. Guinea, N. M. R. Peres, K. S. Novoselov and
A. K. Geim, Rev. Mod. Phys. \textbf{81}, 109 (2009).


\bibitem{tombros07} 
N. Tombros, C. Jozsa, M. Popinciuc, H. T. Jonkman and B. J. van Wees, Nature {\bf 448}, 571 (2007).


\bibitem{sols07}
F. Sols, F. Guinea, and A. H. Castro Neto, Phys. Rev. Lett. {\bf 99}, 166803 (2007).

\bibitem{ponomarenko08}
L. A. Ponomarenko, F. Schedin, M. I. Katsnelson, R. Yang, E. W. Hill, K. S. Novoselov, A. K. Geim, Science {\bf 320}, 356 (2008).

\bibitem{stampfer08}
C. Stampfer, J. Guttinger, F. Molitor, D. Graf, T. Ihn, and K. Ensslin, Appl. Phys. Lett. {\bf 92}, 012102 (2008).

\bibitem{todd09}
K. Todd, H.-T. Chou, S. Amasha, and D. Goldhaber-Gordon, Nano
Letters {\bf 9}, 416 (2009).

\bibitem{Fujita96}
M. Fujita, K. Wakabayashi, K. Nakada, and K. Kusakabe, J. Phys.
Soc. Jpn. \textbf{65}, 1920 (1996).

\bibitem{Yazyev10}
O. V. Yazyev, Rep. Prog. Phys. \textbf{73}, 056501 (2010).

\bibitem{Acik11}
M. Acik and Y. J. Chabal, Jap. J. Appl. Phys. \textbf{50}, 070101
(2011).

\bibitem{Klusek00}
Z. Klusek, Z. Waqar, E. Denisov, T. Kompaniets, I. Makarenko, A.
Titkov, and A. Bhatti, Appl. Surf. Sci. \textbf{161}, 508 (2000).

\bibitem{Kobayashi05}
Y. Kobayashi, K.-i. Fukui, T. Enoki, K. Kusakabe, and Y. Kaburagi,
Phys. Rev. B {\bf 71}, 193406 (2005).

\bibitem{Sutter09}
P. Sutter, Nature Mater. \textbf{8}, 171 (2009).

\bibitem{Son06}
Y.-W. Son, M. L. Cohen, and S. G. Louie, Nature (London)
\textbf{444}, 347 (2006).

\bibitem{Kim08}
W.Y. Kim and K. S. Kim, Nature Nanotech. {\bf 3}, 408 (2008).

\bibitem{Munoz-Rojas09}
F. Mu\~noz-Rojas, J. Fern\'andez-Rossier, and J. J. Palacios,
Phys. Rev. Lett. {\bf 102}, 136810 (2009).

\bibitem{Han11}
W. Han, K.M.McCreary, K. Pi, W.H.Wang, Y. Li, H. Wen, J.R. Chen,
R.K. Kawakami, J. Mag. Mag. Mat. \textbf{324}, 369 (2011).

\bibitem{SK09}
S. Krompiewski, Phys. Rev. B \textbf{80}, 075433 (2009).

\bibitem{Weymann08}
I. Weymann, J. Barna{\'s}, and S. Krompiewski, Phys. Rev. B {\bf 76}, 155408 (2007);
\textbf{78}, 035422 (2008).

\bibitem{Zhou10}
B. Zhou, X. Chen, H. Wang, K. -H. Ding, G. Zhou, J.
Phys.: Condens. Matter \textbf{22}, 445302 (2010).

\bibitem{Tao11}
C. Tao, L. Jiao, O. V. Yazyev, Y.-C. Chen, J. Feng, X. Zhang, R.
B. Capaz, J. M. Tour, A. Zettl, S. G. Louie, H. Dai, and M. F.
Crommie, Nat. Phys. \textbf{7}, 616 (2011).

\bibitem{Areshkin12}
D. A. Areshkin and B. K. Nikoli\'c, Phys. Rev. B {\bf 79}, 205430
(2009).

\bibitem{MZ}
M. Zwierzycki and S. Krompiewski, Acta Physica Polonica A \textbf{118}, 856 (2010).

\bibitem{SK} S. Krompiewski, Nanotechnology, \textbf{22}, 445201 (2011); ibidem, \textbf{23}, 135203 (2012).

\bibitem{Fernandez07}
J. Fern\'andez-Rossier and J. J. Palacios, Phys. Rev. Lett. {\bf
99}, 177204 (2007).

\bibitem{Potasz12}
P. Potasz, A. D. G\"u\c{c}\"u, A. W\'ojs, and P. Hawrylak,
Phys. Rev. B {\bf 85}, 075431 (2012).

\bibitem{Ma09}
Q.~Ma, T.~Tu, Z.-R.~Lin, G.-C.~Guo, G.-P. Guo, cond-mat/0911.2845v1 (2009).

\bibitem{diagrams}
H. Schoeller and G. Sch\"on, Phys. Rev. B {\bf 50}, 18436 (1994);
J. K\"onig, J. Schmid, H. Schoeller, and G. Sch\"on, Phys. Rev. B
{\bf 54}, 16820 (1996).

\bibitem{thielmann}
A. Thielmann, M. H. Hettler, J. K\"onig, and G. Sch\"on, Phys.
Rev. Lett. {\bf 95}, 146806 (2005); Phys. Rev. B {\bf 68}, 115105
(2003).

\bibitem{weymannPRB08}
I. Weymann, Phys. Rev. B {\bf 78}, 045310 (2008).

\bibitem{cotunneling}
D. V. Averin and Yu. V. Nazarov, Phys. Rev. Lett. {\bf 65}, 2446
(1990); K. Kang and B. I. Min, Phys. Rev. B {\bf 55}, 15412
(1997).


\bibitem{Maassen11}
J. Maassen, W. Ji, and H. Guo, Nano Lett. \textbf{11}, 151 (2011).


\bibitem{blanterPR00} 
Ya. M. Blanter and M. B\"uttiker, Phys. Rep. {\bf 336}, 1 (2000).

\bibitem{weymannPRB05}
I. Weymann, J. K\"onig, J. Martinek, J. Barna\'s, and G. Sch\"on,
Phys. Rev. B {\bf 72}, 115334 (2005).


\bibitem{julliere}
M. Julliere, Phys. Lett. A {\bf 54}, 225 (1975).


\bibitem{sukhorukovPRB01}  
E. V. Sukhorukov, G. Burkard, and D. Loss, Phys. Rev. B {\bf 63},
125315 (2001).

\bibitem{onac06}
E. Onac, F. Balestro, B. Trauzettel, C. F. J. Lodewijk, and L. P.
Kouwenhoven, Phys. Rev. Lett. {\bf 96}, 026803 (2006).

\bibitem{zhang}
Y. Zhang, L. DiCarlo, D. T. McClure, M. Yamamoto, S. Tarucha, C.
M. Marcus, M. P. Hanson, and A. C. Gossard,
Phys. Rev. Lett. {\bf 97}, 036603 (2007).

\bibitem{lieb89}
E. H. Lieb, Phys. Rev. Lett. {\bf 62}, 1201 (1989).

\bibitem{Libisch09}
F. Libisch, C. Stampfer, and J. Burgd\"orfer, Phy. Rev. B {\bf 79},
115423 (2009).

\bibitem{Wimmer08}
M. Wimmer, I. Adagideli, S. Berber, D. Tom\'anek, and K. Richter,
Phys Rev. Lett. {\bf 100}, 177207 (2008).


\end{thebibliography}
\end{document}